\documentclass{article}
\usepackage{amsmath,graphicx,mathtools,amssymb,comment}
\usepackage[preprint]{spconf}
\usepackage{array}
\usepackage{booktabs}
\usepackage{colortbl}
\usepackage{multirow}
\usepackage{float}
\usepackage{caption}
\usepackage{xspace}
\usepackage[dvipsnames]{xcolor}
\usepackage[numbers,sort&compress]{natbib}
\usepackage{hyperref}
\usepackage{hhline}
\hypersetup{colorlinks=True}

\def\L{{\cal L}}

\makeatletter
\DeclareRobustCommand\onedot{\futurelet\@let@token\@onedot}
\def\@onedot{\ifx\@let@token.\else.\null\fi\xspace}

\makeatother

\title{LEARNING WORD-LEVEL CONFIDENCE FOR SUBWORD END-TO-END ASR}
\name{\begin{tabular}{c}
      David Qiu$^1$, Qiujia Li$^{2*}$, Yanzhang He$^1$, Yu Zhang$^1$, Bo Li$^1$, Liangliang Cao$^1$,\\
     Rohit Prabhavalkar$^1$, Deepti Bhatia$^1$, Wei Li$^1$, Ke Hu$^1$, Tara N. Sainath$^1$, Ian McGraw$^1$\thanks{$^*$Work was done while the author interned at Google.}
\end{tabular}}
\address{$^1$ Google, LLC, USA, $^2$ University of Cambridge, UK\\
\footnotesize{$^1$\texttt{\{qdavid,yanzhanghe\}@google.com}, $^2$\texttt{ql264@cam.ac.uk}}\vspace{-1em}}
\begin{document}
\ninept

\maketitle
\begin{abstract}
We study the problem of word-level confidence estimation in subword-based end-to-end (E2E) models for automatic speech recognition (ASR). Although prior works have proposed training auxiliary confidence models for ASR systems, they do not extend naturally to systems that operate on word-pieces (WP) as their vocabulary. In particular, ground truth WP correctness labels are needed for training confidence models, but the non-unique tokenization from word to WP causes inaccurate labels to be generated. This paper proposes and studies two confidence models of increasing complexity to solve this problem. The final model uses self-attention to directly learn word-level confidence without needing subword tokenization, and exploits full context features from multiple hypotheses to improve confidence accuracy. Experiments on Voice Search and long-tail test sets show standard metrics (e.g., NCE, AUC, RMSE) improving substantially. The proposed confidence module also enables a model selection approach to combine an on-device E2E model with a hybrid model on the server to address the rare word recognition problem for the E2E model.
\end{abstract}
\section{Introduction}
\label{sec:intro}
\copyrightnotice{\copyright\ IEEE 2021}
\toappear{To appear in {\it Proc.\ ICASSP2021,
                    June 06-11, 2021, Toronto, Ontario, Canada}}
Confidence scores are an important feature of automatic speech recognition (ASR) systems that supports many downstream applications to mitigate ASR errors~\cite{wessel2001confidence, jiang2005confidence, yu2011calibration}. For example, unlabelled utterances with high confidence on the ASR output can be included for semi-supervised learning~\cite{chan2004improving, tur2005combining, park2020improved}.
Words with low word-level confidence can be sent for user correction in spoken dialog systems~\cite{tur2005combining}.
An utterance that has low confidence can be further processed by a different recognizer for improvement.
System combination also commonly relies on confidence as an indication of uncertainty~\cite{Evermann2000PosteriorPD, fiscus1997post}.

In conventional HMM-based hybrid systems, confidence scores are estimated for each output word in the hypotheses. An utterance-level confidence is typically aggregated from the word-level confidence when needed. In such systems, word-level confidence scores can be easily estimated from word posterior probabilities computed from lattices or confusion networks~\cite{evermann2000large,Mangu2000FindingCI}. The estimation can be further improved by model-based approaches to combine word posterior probabilities with optional acoustic, linguistic and duration features using a linear regression model~\cite{evermann2000large, jiang2005confidence}, or more recently, using conditional random fields~\cite{seigel2011combining}, recurrent neural networks~\cite{kalgaonkar2015estimating, ragni2018confidence, swarup2019improving} or graph neural networks~\cite{li2019bi}.

Recently, end-to-end (E2E)  ASR models such as the recurrent neural network transducer (RNN-T)~\cite{ he2019streaming, sainath2020streaming, li2020developing}, transformer or conformer transducer~\cite{zhang2020transformer, yeh2019transformer, gulati2020conformer}, attention-based encoder-decoder models~\cite{chorowski2015attention} \textit{inter alia} have gained popularity and achieved state-of-the-art performance in accuracy and latency~\cite{sainath2020streaming, li2020developing, li2020parallel}. In contrast to conventional hybrid systems, they jointly learn acoustic and language modeling in a single neural network that is E2E trained from data. However, deep neural networks tend to exhibit overconfidence in the prediction~\cite{guo2017calibration, ovadia2019can}. Changes to teacher-forcing maximum likelihood training such as label smoothing~\cite{chorowski2017towards} and scheduled sampling~\cite{bengio2015scheduled} can make the output probabilities less peaky, but the values still do not correlate with word accuracy well.
% are still not reflective of the word error rate (WER).
In our recent work~\cite{li2020confidence}, we quantified the impact of these methods and proposed a confidence estimation module (CEM) that directly learns the correctness label for each hypothesized subword using a binary cross-entropy loss with features from the encoder and decoder of the E2E model. Although the CEM is simple and effective, it learns subword confidence scores (word-pieces in~\cite{li2020confidence}) by using a fixed subword tokenization for each word in the reference sequence, while the hypothesis may contain other valid tokenizations (see Table~\ref{Table: Edit}). This leads to incorrect ground truth labels for training the CEM. 

This paper makes the following contributions: 
1) propose using self-attention in the CEM to learn the word-level confidence directly for a subword ASR without needing subword tokenization, and 2) leverage cross-attention that attends to both acoustic and linguistic context from multiple hypotheses~\cite{hu2020deliberation} for additional gains. Experiments show confidence metrics improving substantially from the baseline CEM that learns subword-level confidence~\cite{li2020confidence} to 1) to 2). For application, we test a confidence-based model selection approach: Each utterance is first recognized by an E2E ASR with the proposed word-level CEM on mobile devices for latency and reliability purposes. If the utterance confidence estimated by the CEM is lower than a pre-set threshold, the utterance is sent to the server to be recognized by a conventional hybrid ASR with a large language model (LM) instead for potential quality improvement. Although the on-device E2E model is more accurate overall on a Voice Search test set than the server hybrid model (5.2\% vs. 6.4\% word error rate (WER)), its lack of LM training data and lexicon causes it to suffer on a rare word test set (17.9\% vs. 9.7\% WER). The model selection approach achieves the best WER on both sets by applying the same threshold on the CEM output (5.2\% on VS and 9.6\% on rare word).

\section{Confidence for Two-Pass ASR}
\label{sec:model}

The base ASR model uses a state-of-the-art two-pass E2E architecture~\cite{sainath2020streaming} introduced in~\cite{li2020parallel}, where the first pass RNN-T generates four candidates for the second pass transformer decoder~\cite{vaswani2017attention} to rerank. Such architecture is proven to achieve low latency and high accuracy streaming recognition on mobile devices. We aim to add a light-weight CEM while maintaining the efficiency of the model.
\begin{figure}[t]
%\vspace{-.1in}
  \centering
    \includegraphics[width=0.45\textwidth]{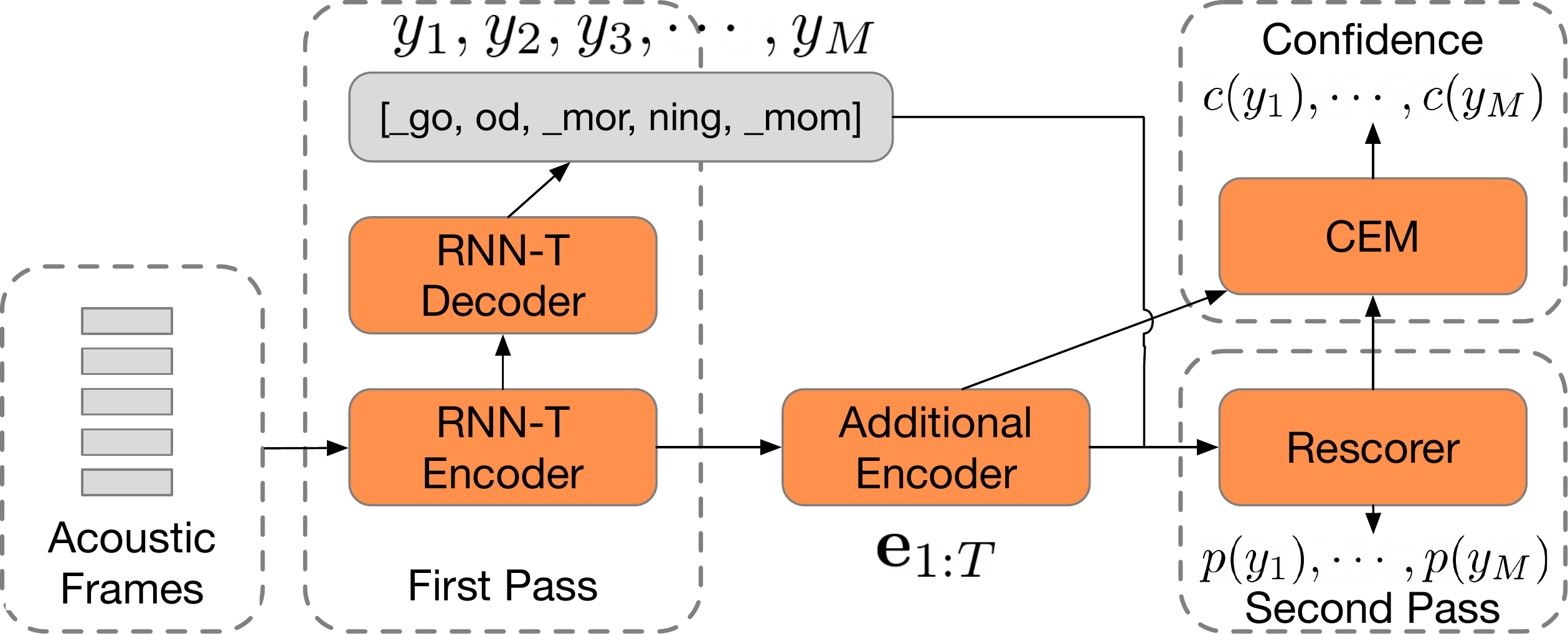}
    \caption{System diagram for the two-pass ASR with confidence.}
    \label{Fig: Two pass}
    \vspace{-.2in}
\end{figure}

For the second pass, the RNN-T is treated as a black box that generates the sequences of acoustic encodings $\mathbf{e} \triangleq \mathbf{e}_{1:T}$ and the hypothesized subword sequence $y_{1:M}$. Fig.~\ref{Fig: Two pass} shows the overall architecture. The rescorer scores each subword using
\begin{align}
    p(y_i|\mathbf{e}, y_{1:i-1}) = \mathrm{Softmax}(\mathrm{Linear}(\phi(i | \mathbf{e}, y_{1:i-1}))),
    \label{Eq: Softmax}
\end{align}
where $\phi$ is the rescorer's penultimate layer activations. The sequence with the highest second pass log probability $\sum_{i=1}^M \log(p(y_i|\mathbf{e}, y_{1:i-1}))$
is output as the transcription. $p(y_i|\mathbf{e}, y_{1:i-1})$ serves as a naive estimate of subword confidence. Until the end of Sec.~\ref{subsec:wpm}, the dependence on $(\mathbf{e}, y_{1:i-1})$ is active even when unwritten.

A dedicated confidence output $c$ can be computed as
\begin{align}
    \textrm{top-}K(i) &:= K \text{ largest log probabilities at decoder index } i \label{Eq: TopN} \\
    b(y_i) &= \left[\mathrm{Emb}(y_i); \phi(i | \mathbf{e}, y_{1:i-1}); \log(p(y_i)); \textrm{top-}K(i) \right] \label{Eq: Features} \\
    c(y_i) &= \sigma(\mathrm{MLP}(b(y_i))). \label{Eq: MLP}
\end{align}
For CEM proposed in \cite{li2020confidence}, a fully-connected multilayer perception (MLP) is used. ``Emb'' is the input subword and position embedding. The CEM can be trained jointly or separately with the ASR, using a binary cross-entropy loss: $\mathcal{L} = -\sum_{i=1}^M d(y_i) \log c(y_i) +  (1-d(y_i))\log(1-c(y_i))$,
where $d(y_i) = 1$ if the edit distance between hypothesized and reference subword sequences outputs ``correct'' for $y_i$ and $d(y_i) = 0$ if it outputs ``insertion'' or ``substitution''.

It is worth noting that the features above that are related to the posterior probability are also commonly used in confidence models for conventional ASR~\cite{evermann2000large, seigel2011combining}. We use them as the baseline for the E2E ASR, but also integrate acoustic and linguistic context with self-attention (Sec.~\ref{subsec:wpm}) and deliberation (Sec.~\ref{subsec:delib}) in a unified network, which dramatically improves the performance over the baseline.

\vspace{-.5em}
\subsection{Simple Word Confidence From Word-pieces}
\label{subsec:simple}
 
To decrease the size of the softmax layer and to improve generalization, the subword vocabulary is typically small compared to the word vocabulary. This can be accomplished with graphemes, word-pieces (WP), etc. In this paper, we focus on WP and use it synonymously with ``subword''. To compute the WER, the hypothesized WP sequence $y_{1:M}$ first needs to be converted to its corresponding word sequence $w_{1:L}$. This procedure is uniquely determined since each word's first WP starts with a word boundary indicator (`\_'), e.g. ``\_go'', ``od'', ``\_mor'', ``ning'' $\to$ ``good'', ``morning''. Similarly for confidence, for a word $w_j$ consisting of $Q_j$ WPs, let $y_{j,q}$ denote the $q$-th WP of the $j$-th word; a simple way to compute word confidence is $c_{\text{w}}(w_j) = \mathrm{agg}(c(y_{j,1}), \ldots, c(y_{j,Q_j}))$, where $\mathrm{agg}$ can be the arithmetic mean, minimum, product, a neural network, etc. In this paper, we experiment with the arithmetic mean aggregator only.

\vspace{-.5em}
\subsection{E2E Word Confidence From Word-pieces}
\label{subsec:wpm}
The drawback of the approach in Sec.~\ref{subsec:simple} lies in the mismatch between WP correctness and word correctness. Even though WP sequences uniquely determine word sequences, the reverse does not hold. Each reference word can be divided into WPs in multiple valid ways. Table~\ref{Table: Edit} shows an example where the word ``morning'' is correctly transcribed, but results in two substitutions in the WP edit distance output. 
\begin{table}[t]
\centering
\begin{tabular}{ l@{\hspace{.3\tabcolsep}}c@{\hspace{0.9\tabcolsep}}c@{\hspace{.3\tabcolsep}}c@{\hspace{0.5\tabcolsep}}c@{\hspace{1.0\tabcolsep}}c } 
 \toprule
 Hyp: & \_go & od & \_mor & ning & \_mom \\ 
 %\hline
 Ref: & \_go & od & \_morn & ing & \\ 
 %\hline
 WP edit: & \textit{cor} & \textit{cor} & \textit{sub} & \textit{sub} & \textit{ins} \\ 
 %\hline
 Word edit: & -- & \textit{cor} & -- & \textit{cor} & \textit{ins} \\
 \midrule
 $d(w_j)$: & -- & $1$ & -- & $1$ & $0$ \\ 
    $m(y_i)$: & $0$ & $1$ & $0$ & $1$ & $1$ \\
    % $\L(y_i)$: & $0$ & $\log c(y_2)$ & $0$ & $\log c(y_4)$ & $\log(1-c(y_5))$ \\
    $\L(w_j)$: & -- & $\log c_{\text{w}}(w_1)$ & -- & $\log c_{\text{w}}(w_2)$ & $\log(1-c_{\text{w}}(w_3))$\\
 \bottomrule
\end{tabular}
\caption{Top: example of a reference having non-unique tokenizations that leads to the inaccurate ground truth WP correctness labels. Bottom: example of using an end-of-word mask $m$ to implement the word-level loss in a CEM with an output at every WP.}
\label{Table: Edit}
\vspace{-.2in}
\end{table}
Searching over all possible reference tokenizations for the one with the fewest WP edits creates an undesirable computational burden during training, and we do not investigate that solution in this paper. Stochastic methods such as BPE-dropout~\cite{provilkov-etal-2020-bpe} also do not help here, since they assume that any segmentation is equally valid, which does not hold when computing edit distance.

Using word edit distance output as the ground truth training labels would bypass the multiple tokenization problem. However, because ASR / CEM output at the WP level, two design choices need to be made: at which WP to output the word confidence, and how to incorporate information from every WP that makes up the word. We choose to use the confidence output at the final WP of every word as its word confidence, and change the MLP in \eqref{Eq: MLP} to a transformer: 
\begin{align}
    \mathbf{b} &= \{b(y_1), \ldots, b(y_{i-1})\} \\
    c(y_i) &= \sigma(\mathrm{Transformer}(\mathrm{CA}(\mathbf{e}), \mathrm{SA}(\mathbf{b}))) \label{Eq: Transformer} \\
    c_{\text{w}}(w_j) &= c(y_{j,Q_j}),
\end{align}
where $\mathrm{CA}$ and $\mathrm{SA}$ denote the cross-attention and self-attention mechanisms~\cite{vaswani2017attention}, respectively, and $b(y_i)$ is defined in \eqref{Eq: Features}. This allows the model to learn how to attend to the features of earlier WPs in the same word in a true E2E fashion. The cross-attention also improves the confidence estimation using the acoustic context.

The word-level loss function becomes 
\vspace{-1em}
\begin{equation*}
    \vspace{-1em}
    L = -\sum_{j=1}^L d(w_j) \log c_{\text{w}}(w_j) +  (1-d(w_j))\log(1-c_{\text{w}}(w_j))
\end{equation*}
Similar to $d(y_i)$, the value of $d(w_j)$ depends on the word edit distance output. The loss, in a WP-based ASR model, can be easily implemented with an end-of-word masked loss (see Table~\ref{Table: Edit}).

\vspace{-.5em}
\subsection{Multi-hypotheses Deliberation}
\label{subsec:delib}
Works such as~\cite{itoh2012n} %gillick1997probabilistic
have shown that statistics from multiple beam search decoded hypotheses further improve confidence accuracy. In general, words shared across more hypotheses tend to have higher confidence. Information from multiple hypotheses is even more relevant to the model introduced in Sec.~\ref{subsec:wpm}. In the example from Table~\ref{Table: Edit}, having two hypotheses [\_go, od, \_mor, ning, \_mom] and [\_go, od, \_morn, ing, \_mom] attend to each other would inform the model that they concatenate to the same word sequence, and that they should be mapped to similar confidence scores.
Additionally, the goal of confidence prediction is to score a known hypothesis. This is different from auto-regressive decoding, where knowing the full hypothesis trivializes the problem. Thus, the CEM can make use of the future context of the hypothesis to score the current word.

\begin{figure}[t]
% \vspace{-.1in}
  \centering
    \includegraphics[width=0.94\linewidth]{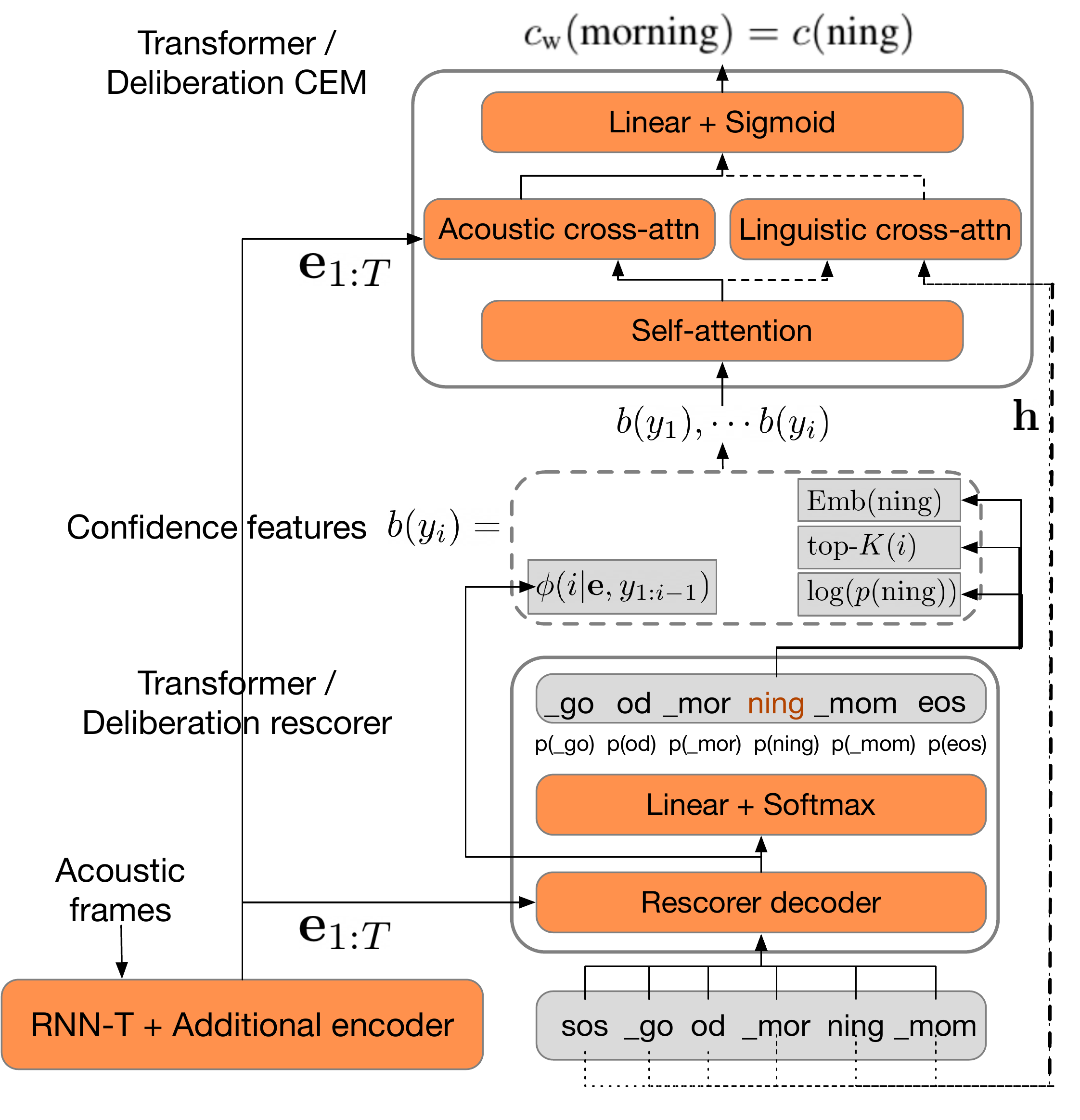}
    \vspace{-.1in}
    \caption{Model architecture. For clarity, only the actions of predicting $c(\text{``ning''})$ are shown. Refer to Table~\ref{Table: Edit} for an example of the masked loss function for the entire sequence. Top-$K$ is defined in \eqref{Eq: TopN}. All dashed connections and the linguistic cross-attention block are only used in the deliberation CEM but not transformer CEM.}
    \label{Fig: Architecture}
    \vspace{-.25in}
\end{figure}

We incorporate two sources of information, acoustic encoding ($\mathbf{e}$) and multiple hypotheses encoding ($\mathbf{h}$), in a learned way through the multi-source attention block in a deliberation model~\cite{hu2020deliberation}:
\begin{align}
    \mathbf{h} &= \left[\mathrm{BLSTM}\left(y_{1:M_1}^{(1)} \right); \ldots ; \mathrm{BLSTM}\left(y_{1:M_H}^{(H)} \right)\right] \\
    c(y_i) &= \sigma(\mathrm{Transformer}(\mathrm{CA}(\mathbf{e}) + \mathrm{CA}(\mathbf{h}), \mathrm{SA}(\mathbf{b}))),
\end{align}
where $H$ is the number of hypotheses attended to and $M_H$ is the number of WPs in the $H$-th hypothesis. Fig.~\ref{Fig: Architecture} shows the model.

\vspace{-.3em}
\section{Experimental Setup}
\label{sec:experiments}

% \vspace{-.1in}
% \subsection{Architecture and Training Setup}
\textbf{Architecture and Training Setup:} The RNN-T architecture follows~\cite{sainath2020streaming} exactly, with 8 LSTM layers in the encoder and 2 LSTM layers in the prediction network. Each LSTM layer is unidirectional, with 2,048 units and a projection layer with 640 units. The transformer rescorer architecture is the same as in~\cite{li2020parallel}. Its encoder consists of the shared encoder from RNN-T and additional 2 LSTM layers, and the decoder consists of 4 self-attention layers, 2 of which contain the cross-attention over the encoder. The deliberation rescorer architecture is described in~\cite{hu2020deliberation}; for consistency, we replace its LAS decoder with the same 4 layers transformer decoder. We feed up to eight RNN-T beam search results into the linguistic cross-attention mechanism. All internal dimensions in the rescorers are 640. The size of the WP vocabulary is 4,096. The ASR model is frozen during CEM training to not affect the WER. 

For the top-$K$ feature in \eqref{Eq: TopN}, we observe diminishing returns beyond $K = 4$, and use this setting for all experiments. Thus, the input features introduced in \eqref{Eq: Features} are 640, 640, 1, 4 dimensional, respectively. ``WP MLP'' denotes the CEM in~\eqref{Eq: MLP} with confidence averaged at the word level (Sec.~\ref{subsec:simple}); we use 3 layers that result in hidden activation with dimensions of 640, 320, 1, respectively. We also replace the second layer with one transformer decoder block and call this setup ``WP Xformer''. ``E2E Xformer'' denotes the CEM in Sec.~\ref{subsec:wpm}. All confidence models are trained in TensorFlow with the Lingvo~\cite{shen2019lingvo} toolkit using four hypotheses from the frozen RNN-T, and evaluated on only the top reranked hypothesis. The optimizer is Adam~\cite{KingmaB14} with learning rate 0.0005, and the global batch size is 4,096 across $8 \times 8$ TPU. %All models are evaluated after approximately 200,000 steps.

% \vspace{-.1in}
% \subsection{Data}
\textbf{Training Set:} The models are trained on the multi-domain training set used in~\cite{sainath2020streaming}, which spans domains of search, farfield, telephony and YouTube. All datasets are anonymized and hand-transcribed; the transcription for YouTube utterances is done in a semi-supervised fashion~\cite{liao2013large}. Multi-condition training (MTR)~\cite{kim2017generation} and random data downsampling to 8kHz~\cite{li2012improving} are used to further increase data diversity.

\textbf{Test Sets:} The main test set includes $\sim$14K Voice Search (VS) utterances extracted from Google traffic, which is anonymized and hand-transcribed. To test the generalizability of the CEM,
we use an in-house named entity tagger to identify a list of proper nouns that are common in the LM training data for conventional ASR but rare in the audio-text paired multi-domain training data for E2E ASR models. We select 10,000 sentences from the LM test data for the maps domain, each of which contains at least one of these rare proper nouns, then synthesize audio for these sentences with a TTS system (as in~\cite{gonzalvo2016recent}) to create the Long-tail Maps test set.
\begin{table*}[t]
    \centering
    \begin{tabular}{l|ccc|cc|ccc|cc}
        \toprule
          & \multicolumn{5}{c|}{Voice Search} & \multicolumn{5}{c}{Long-tail Maps} \\
          \cmidrule{2-11}
        Confidence Models  &  \multirow{2}{*}{NCE} & AUC & AUC & WCR & (1-WER) & \multirow{2}{*}{NCE} & AUC & AUC & WCR & (1-WER) \\
          & & ROC & PR  & RMSE & RMSE & & ROC & PR  & RMSE & RMSE \\
        \midrule
        ASR Softmax & 0.241 & 0.873 & 0.280 & 0.140 & 0.244 & 0.286 & 0.882 & 0.635 & 0.221 & 0.334\\
          WP MLP~\cite{li2020confidence} & 0.269 & 0.885	& 0.329	& 0.138 & 0.233 & 0.360 & 0.887 & 0.684 & 0.198	& 0.297 \\ \hhline{===========}
          WP Xformer & 0.280	& 0.885	& 0.347	& 0.137	& 0.231 & 0.365 & 0.889 & 0.690 & 0.194 & 0.292 \\
        \hline
        E2E Xformer & 0.367 & 0.928 & 0.466 & 0.130 & 0.221 & 0.389 & 0.901 & 0.682 & 0.186 & \textbf{0.281} \\
          $+$Delib 1-Hyp & 0.361 & 0.923 & 0.474 & 0.128 &	0.206 & 0.405 & 0.908 & 0.691 & \textbf{0.184} & 0.299 \\
          $+$Delib 8-Hyp & \textbf{0.425} & \textbf{0.941} & \textbf{0.508} & \textbf{0.127} & \textbf{0.204} & \textbf{0.416} & \textbf{0.911} & \textbf{0.700} & \textbf{0.184} & 0.283 \\
        \bottomrule
    \end{tabular}
    \caption{Confidence metrics comparing the WP, E2E, and deliberation CEM. The models in the last 4 rows are proposed in this paper.}
    \label{Table: VS}
    % \vspace{-.1in}
\end{table*}

% \vspace{-.1in}
% \subsection{Evaluation Metrics}
\textbf{Evaluation Metrics:} We evaluate on standard metrics found in prior works in confidence~\cite{ragni2018confidence}. To measure per-word confidence accuracy, we use normalized cross-entropy (NCE)~\cite{Siu1997ImprovedEE}. To measure whether the confidence score is highly correlated with word correctness, we use area under the receiver operating characteristic curve (AUC-ROC) and the precision recall curve (AUC-PR). Because the ASR model achieves high word correct ratio
$\left( \text{WCR} = \frac{\text{\#(correctly hypothesized words)}}{\text{\#(all hypothesized words)}} \right)$ on all datasets, we compute AUC-PR with the PR curve for the less frequent incorrect class (wrongly hypothesized words). Higher is better for these metrics.

Utterance-level confidence is computed by averaging word-level confidence: $\frac{1}{L} \sum_{j=1}^L c_{\text{w}}(w_j)$. To measure its accuracy, we use the root mean squared error (RMSE) between the utterance confidence and either the ground truth WCR or $(1-\text{WER})$. Although WER is the gold standard, because this version of the CEM does not explicitly predict deletions, WCR RMSE is a better indicator of the CEM's quality. Lower is better for these metrics.

\vspace{-.5em}
\section{Results}
\label{sec:results}
\vspace{-.7em}
\subsection{Simple vs E2E Word Confidence from Word-pieces}
\begin{figure}[ht]
\vspace{-.2in}
\begin{minipage}[b]{.49\linewidth}
  \centering
  \centerline{\includegraphics[width=4.3cm]{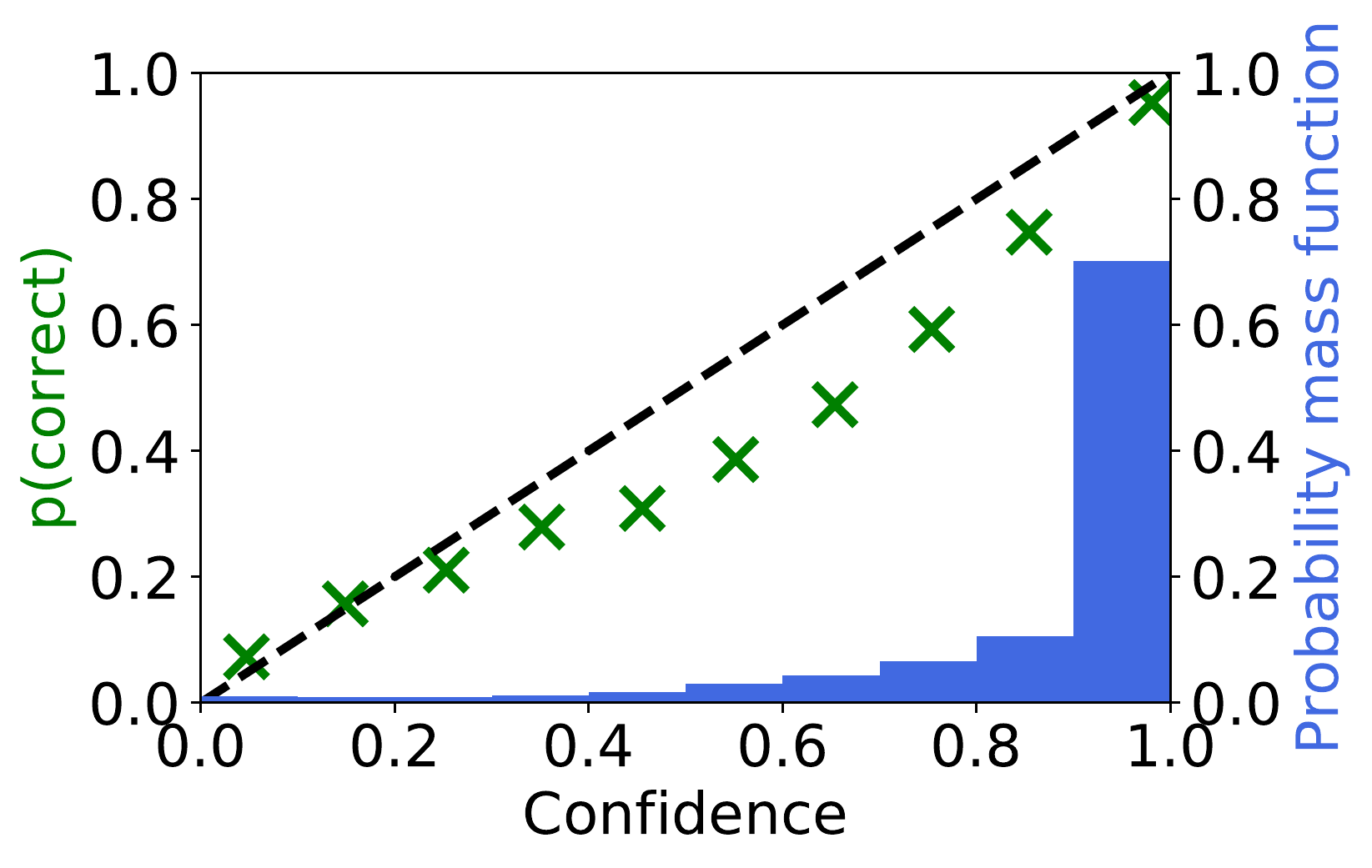}}
%  \vspace{1.5cm}
  \centerline{(a) ASR Softmax}\medskip
\end{minipage}
\hfill
\begin{minipage}[b]{0.49\linewidth}
  \centering
  \centerline{\includegraphics[width=4.3cm]{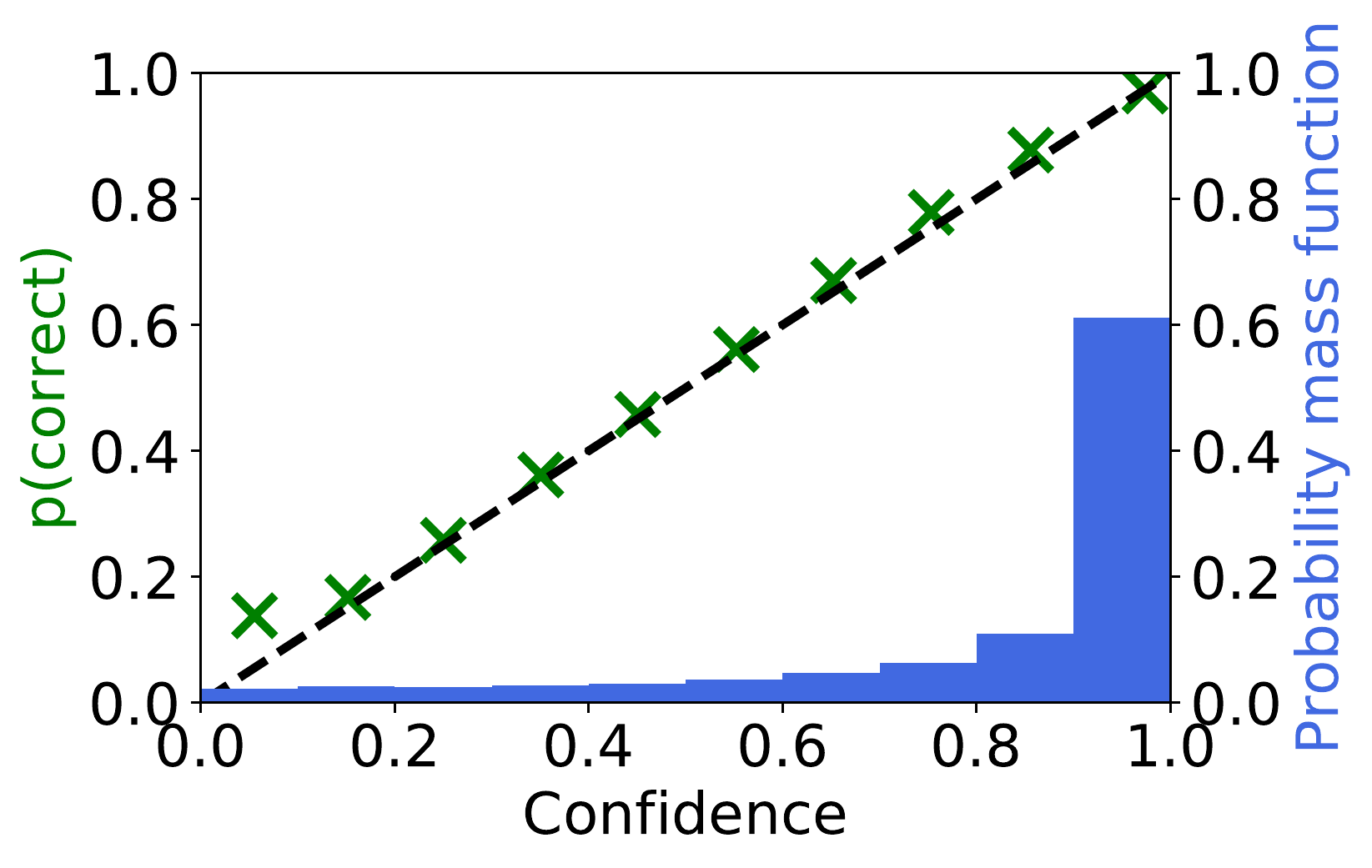}}
%  \vspace{1.5cm}
  \centerline{(b) E2E Xformer}\medskip
\end{minipage}
\vspace{-.1in}
\caption{Calibration curves for ASR Softmax and E2E Xformer confidence models for word confidence on Long-tail Maps. The black and green curves show the ideal and actual calibration curves, respectively. The blue bar plot shows the probability mass in each bin.}
\label{Fig: Calibration}
\vspace{-.2in}
\end{figure}

\vspace{-.2em}
This section compares the performance of the naive confidence from the ASR softmax, WP CEM (Sec.~\ref{subsec:simple}), and E2E CEM (Sec.~\ref{subsec:wpm}). Table~\ref{Table: VS} shows the results of the representative models on VS and Maps. Except AUC-PR on Maps, all other metrics improve going from softmax to WP to E2E, showing the effectiveness of our proposed technique. Deep neural networks usually exhibits overconfidence on long-tail data, and Fig.~\ref{Fig: Calibration} shows the improved calibration curve~\cite{guo2017calibration} on Maps arising from E2E word confidence training.

\vspace{-.7em}
\subsection{Effects of Input Features}
\vspace{-.2em}
To determine the effects of input embedding in \eqref{Eq: Features}, we compare using the ASR input embedding against training a new embedding layer for confidence (``Conf emb''). To quantify the usefulness of the ASR softmax posteriors, we compare using the full set of features in \eqref{Eq: Features}, removing the top-$K$ ($K=4$) feature, and further removing the log posterior $\log(p(y_i))$.
Table~\ref{Table: Ablation} reports the metrics from the input feature ablation study. Having a dedicated confidence embedding layer improves all metrics, at the cost of $4096 \times 640$ extra parameters. The model degrades when it does not use the log probability features. Given the modest increase in dimensionality from these five additional features, it is worth including them in the CEM.
\begin{table}[ht]
\vspace{-.1in}
    \centering
    \begin{tabular}{l|ccc|cc}
        \toprule
          &  \multirow{2}{*}{NCE} & AUC & AUC & WCR & (1-WER) \\
          & & ROC & PR  & RMSE & RMSE \\
        \midrule
        E2E Xformer & 0.367 & 0.928 & 0.466 & 0.130 & 0.221 \\
          $+$Conf emb & 0.374 & 0.930 & 0.477 & 0.129 & 0.219 \\
          $-$top-$K(i)$ & 0.358 & 0.924 & 0.453 & 0.131 & 0.222 \\
          \hspace{2mm}$-\log(p(y_i))$ & 0.338 & 0.924 & 0.435 & 0.136 & 0.223 \\
        \bottomrule
    \end{tabular}
    \caption{Input feature ablation studies on Voice Search.}
    \label{Table: Ablation}
    \vspace{-1.5em}
\end{table}

\vspace{-.5em}
\subsection{Multi-hypotheses Deliberation Results}
This section examines adding multi-hypotheses deliberation to further improve the E2E Xformer CEM. The Delib 1-Hyp model uses a BLSTM to encode the current hypothesis and changes every transformer layer in the ASR and CEM to use multi-source attention that attends to both the acoustic and the BLSTM hypothesis encodings. This allows the CEM to see future context in the hypothesis. The Delib 8-Hyp model uses the RNN-T to generate eight hypotheses to be encoded and attended to. Encodings across different hypotheses are concatenated without any position information. This allows the CEM to use consensus among multiple hypotheses.

\begin{figure}[ht]
\vspace{-.2in}
  \centering
    \includegraphics[width=0.6\linewidth]{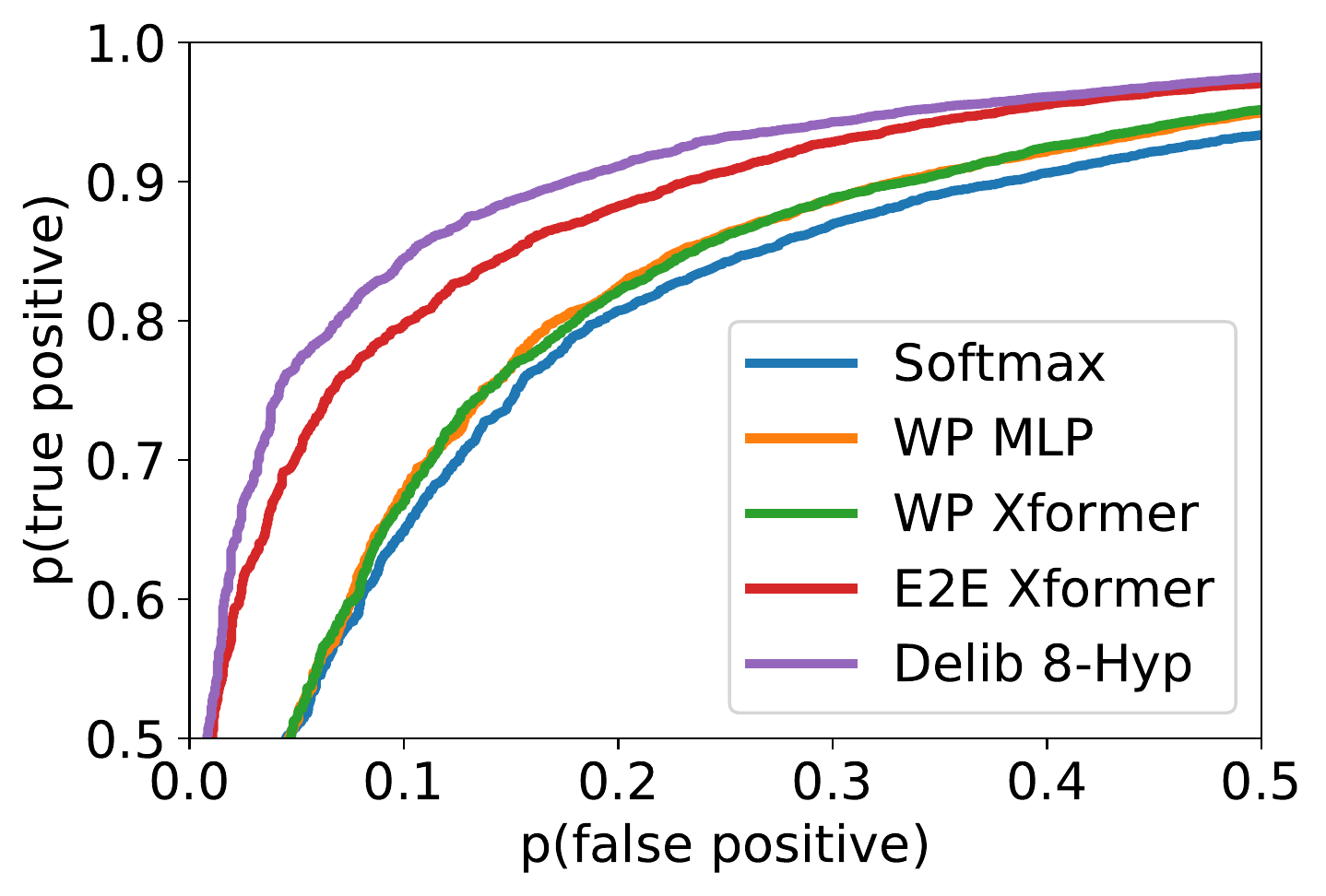}
    \vspace{-.1in}
    \caption{ROC curve on Voice Search for different models.}
    \label{Fig: ROC}
    \vspace{-.2in}
\end{figure}

Table~\ref{Table: VS} shows that using full context on the single hypothesis (Delib 1-Hyp) slightly improves some of the metrics. However, consensus from multiple hypotheses (Delib 8-Hyp) greatly improves most metrics over the basic E2E Xformer model. Fig.~\ref{Fig: ROC} demonstrates that in a ROC curve, where E2E CEM is clearly better than WP CEM, and adding deliberation improves further.

\vspace{-.5em}
\subsection{Long-tail Utterances Filtering for Model Selection}
\begin{figure}[h!]
\vspace{-.1in}
\begin{minipage}[b]{.49\linewidth}
  \centering
  \centerline{\includegraphics[width=4.3cm]{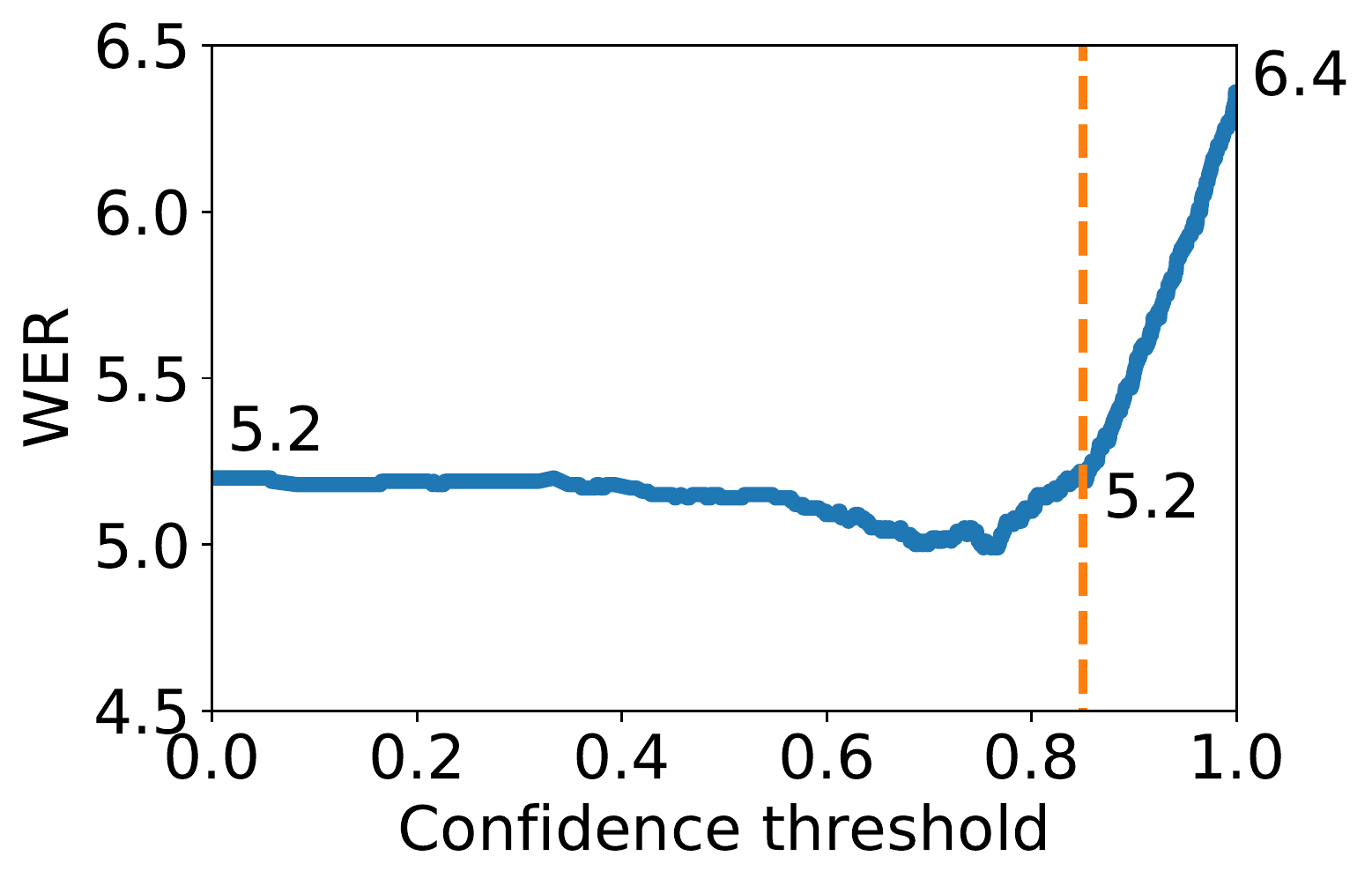}}
%  \vspace{1.5cm}
  \centerline{(a) Voice Search}\medskip
\end{minipage}
\hfill
\begin{minipage}[b]{0.49\linewidth}
  \centering
  \centerline{\includegraphics[width=4.3cm]{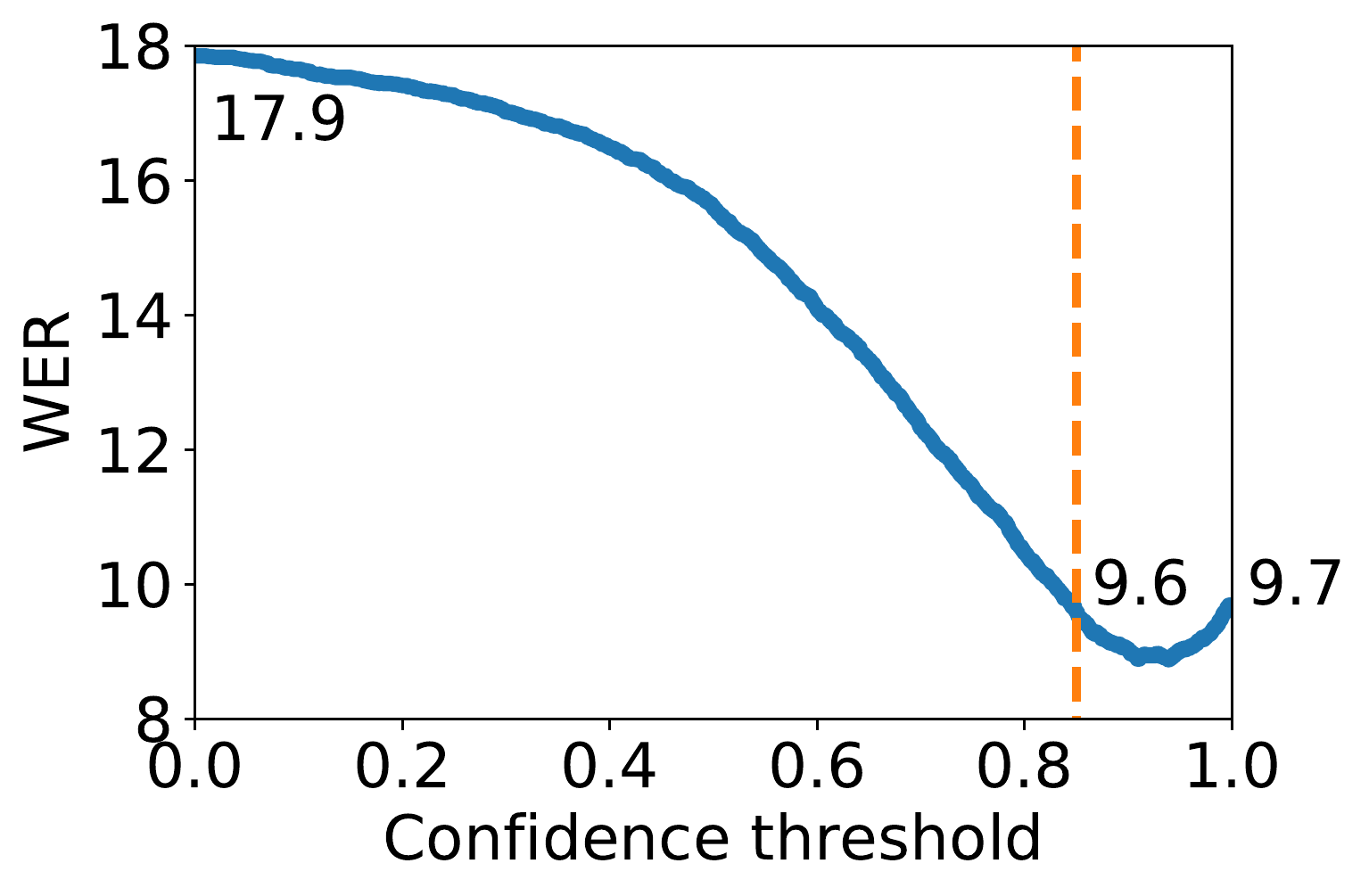}}
%  \vspace{1.5cm}
  \centerline{(b) Long-tail Maps}\medskip
\end{minipage}
\vspace{-.1in}
\caption{Overall WER at different operating points for model selection. When confidence threshold is 0, all utterances are processed on-device. When it is 1, all utterances are processed on the server.}
\label{Fig: Hybrid}
\vspace{-.2in}
\end{figure}

For application, we experiment with a confidence-based model selection approach to combine 1) an on-device two-pass E2E ASR with deliberation rescoring and an E2E Delib 8-Hyp confidence module (5.2\% VS WER), and 2) a conventional hybrid ASR on the server~\cite{pundak2016lower} with a large LM (6.4\% VS WER). Despite the lower WER on VS, the E2E model's lack of large-scale LM training data and lexicon causes it to suffer on long-tail utterances (17.9\% on Maps compared to 9.7\% with server). Thus, the objective is to send all utterances with the confidence score below a pre-set threshold to the server, while keeping the majority of utterances on-device to gain quality, latency, and reliability.  Fig.~\ref{Fig: Hybrid} shows the WER of the overall system with different confidence thresholds. When the threshold is set to 0.85, 87\% of the VS utterances are processed on-device, and the overall WER is equal to on-device only (5.2\%). With the same threshold, only 53\% of the Maps utterances are processed on-device, and the overall WER of 9.6\% is better than either individual system.

\section{Conclusion}
\label{sec:conclusion}
We propose an extension to a light-weight confidence estimation module for E2E ASR models to directly estimate word-level confidence with self-attention and deliberation, by learning from the full acoustic and linguistic context of subword sequence and multiple hypotheses. Experimental results show the proposed approach is critical to improving confidence metrics substantially when applied to a state-of-the-art two-pass E2E system. It also enables a confidence-based model selection approach to address the rare word recognition problem for the E2E system.

% Below is an example of how to insert images. Delete the ``\vspace'' line,
% uncomment the preceding line ``\centerline...'' and replace ``imageX.ps''
% with a suitable PostScript file name.
% -------------------------------------------------------------------------

%\vfill
%\pagebreak

\vfill\pagebreak

\section{References}
\vspace{-0.3em}
\begingroup
\renewcommand{\section}[2]{}
\bibliographystyle{IEEEbib}
\bibliography{refs}

\begin{thebibliography}{10}

\bibitem{wessel2001confidence}
F.~Wessel, R.~Schluter, K.~Macherey, \& H.~Ney,
\newblock ``Confidence measures for large vocabulary continuous speech
  recognition,''
\newblock {\em IEEE Trans. on Speech and Audio Processing}, 2001.

\bibitem{jiang2005confidence}
H.~Jiang,
\newblock ``Confidence measures for speech recognition: A survey,''
\newblock {\em Speech communication}, 2005.

\bibitem{yu2011calibration}
D.~Yu, J.~Li, \& L.~Deng,
\newblock ``Calibration of confidence measures in speech recognition,''
\newblock {\em IEEE Trans. on Audio, Speech, and Language Processing}, 2011.

\bibitem{chan2004improving}
H.Y.~Chan \& P.C.~Woodland,
\newblock ``Improving broadcast news transcription by lightly supervised
  discriminative training,''
\newblock in {\em ICASSP}, 2004.

\bibitem{tur2005combining}
G.~Tur, D.~Hakkani-T{\"u}r, \& R.E.~Schapire,
\newblock ``Combining active and semi-supervised learning for spoken language
  understanding,''
\newblock {\em Speech Communication}, 2005.

\bibitem{park2020improved}
D.~Park, Y.~Zhang, Y.~Jia, W.~Han, C.C.~Chiu, \etal,
\newblock ``Improved noisy student training for automatic speech recognition,''
\newblock in {\em Interspeech}, 2020.

\bibitem{Evermann2000PosteriorPD}
G.~Evermann \& P.C.~Woodland,
\newblock ``Posterior probability decoding, confidence estimation and system
  combination,''
\newblock in {\em NIST Speech Transcription Workshop}, 2000.

\bibitem{fiscus1997post}
J.G.~Fiscus,
\newblock ``A post-processing system to yield reduced word error rates:
  Recognizer output voting error reduction ({ROVER}),''
\newblock in {\em ASRU}, 1997.

\bibitem{evermann2000large}
G.~Evermann \& P.C.~Woodland,
\newblock ``Large vocabulary decoding and confidence estimation using word
  posterior probabilities,''
\newblock in {\em ICASSP}, 2000.

\bibitem{Mangu2000FindingCI}
L.~Mangu, E.~Brill, \& A.~Stolcke,
\newblock ``Finding consensus in speech recognition: Word error minimization
  and other applications of confusion networks,''
\newblock {\em Computer Speech and Language}, 2000.

\bibitem{seigel2011combining}
M.S.~Seigel \& P.C.~Woodland,
\newblock ``Combining information sources for confidence estimation with {CRF}
  models,''
\newblock in {\em Interspeech}, 2011.

\bibitem{kalgaonkar2015estimating}
K.~Kalgaonkar, C.~Liu, Y.~Gong, \& K.~Yao,
\newblock ``Estimating confidence scores on {ASR} results using recurrent
  neural networks,''
\newblock in {\em ICASSP}, 2015.

\bibitem{ragni2018confidence}
A.~Ragni, Q.~Li, M.J.F.~Gales, \& Y.~Wang,
\newblock ``Confidence estimation and deletion prediction using bidirectional
  recurrent neural networks,''
\newblock in {\em SLT}, 2018.

\bibitem{swarup2019improving}
P.~Swarup, R.~Maas, S.~Garimella, S.H.~Mallidi, \& B.~Hoffmeister,
\newblock ``Improving {ASR} confidence scores for alexa using acoustic and
  hypothesis embeddings,''
\newblock {\em Interspeech}, 2019.

\bibitem{li2019bi}
Q.~Li, P.~Ness, A.~Ragni, \& M.J.F.~Gales,
\newblock ``Bi-directional lattice recurrent neural networks for confidence
  estimation,''
\newblock in {\em ICASSP}, 2019.

\bibitem{he2019streaming}
Y.~He, T.~Sainath, R.~Prabhavalkar, I.~McGraw, R.~Alvarez, \etal,
\newblock ``Streaming end-to-end speech recognition for mobile devices,''
\newblock in {\em ICASSP}, 2019.

\bibitem{sainath2020streaming}
T.~Sainath, Y.~He, B.~Li, A.~Narayanan, R.~Pang, \etal,
\newblock ``A streaming on-device end-to-end model surpassing server-side
  conventional model quality and latency,''
\newblock in {\em ICASSP}, 2020.

\bibitem{li2020developing}
J.~Li, R.~Zhao, Z.~Meng, Y.~Liu, W.~Wei, \etal,
\newblock ``Developing {RNN-T} models surpassing high-performance hybrid models
  with customization capability,''
\newblock in {\em Interspeech}, 2020.

\bibitem{zhang2020transformer}
Q.~Zhang, H.~Lu, H.~Sak, A.~Tripathi, E.~McDermott, \etal,
\newblock ``Transformer transducer: A streamable speech recognition model with
  transformer encoders and {RNN-T} loss,''
\newblock in {\em ICASSP}, 2020.

\bibitem{yeh2019transformer}
C.F.~Yeh, J.~Mahadeokar, K.~Kalgaonkar, Y.~Wang, D.~Le, \etal,
\newblock ``Transformer-transducer: End-to-end speech recognition with
  self-attention,''
\newblock {\em arXiv:1910.12977}, 2019.

\bibitem{gulati2020conformer}
A.~Gulati, J.~Qin, C.C.~Chiu, N.~Parmar, Y.~Zhang, \etal,
\newblock ``Conformer: Convolution-augmented transformer for speech
  recognition,''
\newblock in {\em Interspeech}, 2020.

\bibitem{chorowski2015attention}
J.K.~Chorowski, D.~Bahdanau, D.~Serdyuk, \etal,
\newblock ``Attention-based models for speech recognition,''
\newblock in {\em NeurIPS}, 2015.

\bibitem{li2020parallel}
W.~Li, J.~Qin, C.C.~Chiu, R.~Pang, \& Y.~He,
\newblock ``Parallel rescoring with transformer for streaming on-device speech
  recognition,''
\newblock in {\em Interspeech}, 2020.

\bibitem{guo2017calibration}
C.~Guo, G.~Pleiss, Y.~Sun, \& K.Q.~Weinberger,
\newblock ``On calibration of modern neural networks,''
\newblock in {\em ICML}, 2017.

\bibitem{ovadia2019can}
Y.~Ovadia, E.~Fertig, J.~Ren, Z.~Nado, D.~Sculley, \etal,
\newblock ``Can you trust your model's uncertainty? {E}valuating predictive
  uncertainty under dataset shift,''
\newblock in {\em NeurIPS}, 2019.

\bibitem{chorowski2017towards}
J.~Chorowski \& N.~Jaitly,
\newblock ``Towards better decoding and language model integration in sequence
  to sequence models,''
\newblock in {\em Interspeech}, 2017.

\bibitem{bengio2015scheduled}
S.~Bengio, O.~Vinyals, N.~Jaitly, \& N.~Shazeer,
\newblock ``Scheduled sampling for sequence prediction with recurrent neural
  networks,''
\newblock in {\em NIPS}, 2015.

\bibitem{li2020confidence}
Q.~Li, D.~Qiu, Y.~Zhang, B.~Li, Y.~He, \etal,
\newblock ``Confidence estimation for attention-based sequence-to-sequence
  models for speech recognition,''
\newblock {\em arXiv:2010.11428}, 2020.

\bibitem{hu2020deliberation}
K.~Hu, T.N.~Sainath, R.~Pang, \& R.~Prabhavalkar,
\newblock ``Deliberation model based two-pass end-to-end speech recognition,''
\newblock in {\em ICASSP}, 2020.

\bibitem{vaswani2017attention}
A.~Vaswani, N.~Shazeer, N.~Parmar, J.~Uszkoreit, L.~Jones, \etal,
\newblock ``Attention is all you need,''
\newblock in {\em NeurIPS}, 2017.

\bibitem{provilkov-etal-2020-bpe}
I.~Provilkov, D.~Emelianenko, \& E.~Voita,
\newblock ``{BPE-Dropout}: Simple and effective subword regularization,''
\newblock in {\em ACL}, 2020.

\bibitem{itoh2012n}
N.~Itoh, T.N.~Sainath, D.N.~Jiang, J.~Zhou, \& B.~Ramabhadran,
\newblock ``N-best entropy based data selection for acoustic modeling,''
\newblock in {\em ICASSP}, 2012.

\bibitem{shen2019lingvo}
J.~Shen, P.~Nguyen, Y.~Wu, Z.~Chen, M.X.~Chen, \etal,
\newblock ``Lingvo: A modular and scalable framework for sequence-to-sequence
  modeling,''
\newblock {\em arXiv:1902.08295}, 2019.

\bibitem{KingmaB14}
D.P.~Kingma \& J.~Ba,
\newblock ``Adam: {A} method for stochastic optimization,''
\newblock in {\em ICLR}, 2015.

\bibitem{liao2013large}
H.~Liao, E.~McDermott, \& A.~Senior,
\newblock ``Large scale deep neural network acoustic modeling with
  semi-supervised training data for youtube video transcription,''
\newblock in {\em ASRU}, 2013.

\bibitem{kim2017generation}
C.~Kim, A.~Misra, K.K.~Chin, T.~Hughes, A.~Narayanan, \etal,
\newblock ``Generation of large-scale simulated utterances in virtual rooms to
  train deep-neural networks for far-field speech recognition in {Google
  Home},''
\newblock in {\em Interspeech}, 2017.

\bibitem{li2012improving}
J.~Li, D.~Yu, J.T.~Huang, \& Y.~Gong,
\newblock ``Improving wideband speech recognition using mixed-bandwidth
  training data in {CD-DNN-HMM},''
\newblock in {\em SLT}, 2012.

\bibitem{gonzalvo2016recent}
X.~Gonzalvo, S.~Tazari, C.a.~Chan, M.~Becker, A.~Gutkin, \etal,
\newblock ``Recent advances in {Google} real-time {HMM}-driven unit selection
  synthesizer,''
\newblock in {\em Interspeech}, 2016.

\bibitem{Siu1997ImprovedEE}
M.~Siu, H.~Gish, \& F.~Richardson,
\newblock ``Improved estimation, evaluation and applications of confidence
  measures for speech recognition,''
\newblock in {\em Eurospeech}, 1997.

\bibitem{pundak2016lower}
G.~Pundak \& T.N.~Sainath,
\newblock ``Lower frame rate neural network acoustic models,''
\newblock in {\em Interspeech}, 2016.

\end{thebibliography}
\endgroup

\end{document}